\documentclass[11pt]{amsart}

\reversemarginpar
\newcommand{\commentout}[1]{}

\usepackage{amsmath}
\usepackage{amssymb}

\newcommand{\gradx}{\nabla_{\bx}}
\newcommand{\Hbar}{\bar{H}}

\newcommand{\nwc}{\newcommand}

\newcommand{\lt}{\left}
\newcommand{\rt}{\right}

\newcommand{\lan}{\left\langle}
\newcommand{\ran}{\right\rangle}

\newcommand{\bx}{\mathbf x}
\nwc{\xbar}{\bar{\mathbf x}}
\nwc{\pbar}{\bar{\mathbf p}}
\nwc{\zbar}{\bar{z}}
\nwc{\zhat}{\hat{z}}
\nwc{\zm}{{z_m}}
\nwc{\zo}{{z_0}}
\newcommand{\bp}{\mathbf p}
\newcommand{\by}{\mathbf y}

\newcommand{\bq}{\mathbf q}
\newcommand{\bw}{\mathbf w}
\newcommand{\br}{\mathbf r}

\nwc{\nwt}{\newtheorem}

\nwt{proposition}{Proposition}
\nwt{prop}{Proposition}
\nwt{lemma}{Lemma}
\nwt{cor}{Corollary}
\nwt{theorem}{Theorem}
\nwt{remark}{Remark}
\nwt{definition}{Definition} 

\nwc{\bal}{\begin{align}}
\nwc{\be}{\begin{equation}}
\nwc{\ben}{\begin{equation*}}
\nwc{\bea}{\begin{eqnarray}}
\nwc{\beq}{\begin{eqnarray}}
\nwc{\bean}{\begin{eqnarray*}}
\nwc{\beqn}{\begin{eqnarray*}}
\nwc{\beqast}{\begin{eqnarray*}}

\nwc{\eal}{\end{align}}
\nwc{\ee}{\end{equation}}
\nwc{\een}{\end{equation*}}
\nwc{\eea}{\end{eqnarray}}
\nwc{\eeq}{\end{eqnarray}}
\nwc{\eean}{\end{eqnarray*}}
\nwc{\eeqn}{\end{eqnarray*}}
\nwc{\eeqast}{\end{eqnarray*}}

\nwc{\tx}{\tilde{\bx}}
\nwc{\tp}{\tilde{\bp}}
\nwc{\tr}{\tilde{\br}}
\nwc{\tw}{\tilde{\bw}}
\nwc{\ep}{\varepsilon}
\nwc{\ept}{\epsilon}
\nwc{\vrho}{\varrho}
\nwc{\orho}{\bar\varrho}
\nwc{\ou}{\bar u}
\nwc{\vpsi}{\varpsi}
\nwc{\lamb}{\lambda}
\nwc{\wep}{W^\ep}

\nwc{\partz}{\frac{\partial }{\partial z}}
\nwc{\partt}{\frac{\partial }{\partial \tau}}

\nwc{\nn}{\nonumber}
\nwc{\bm}{\boldmath}
\nwc{\mf}{\mathbf}
\nwc{\mb}{\mathbf}
\nwc{\ml}{\mathcal}

\nwc{\bD}{{\mb D}}

\nwc{\IA}{\mathbb{A}} 
\nwc{\IB}{\mathbb{B}}
\nwc{\IC}{\mathbb{C}} 
\nwc{\ID}{\mathbb{D}} 
\nwc{\IM}{\mathbb{M}} 
\nwc{\IP}{\mathbb{P}} 
\nwc{\II}{\mathbb{I}} 
\nwc{\IE}{\mathbb{E}} 
\nwc{\IF}{\mathbb{F}} 
\nwc{\IG}{\mathbb{G}} 
\nwc{\IN}{\mathbb{N}} 
\nwc{\IQ}{\mathbb{Q}} 
\nwc{\IR}{\mathbb{R}} 
\nwc{\IT}{\mathbb{T}} 
\nwc{\IZ}{\mathbb{Z}} 

\nwc{\cE}{{\ml E}}
\nwc{\cP}{{\ml P}}
\nwc{\cL}{{\ml L}}
\nwc{\cN}{{\ml N}}
\nwc{\cU}{{\ml U}}
\nwc{\cR}{{\ml R}}
\nwc{\cV}{{\ml V}}
\nwc{\cW}{{\ml W}}
\nwc{\cT}{{\ml T}}
\nwc{\crV}{{\ml V}_{(\delta,\rho)}}
\nwc{\cC}{{\ml C}}
\nwc{\cA}{{\ml A}}
\nwc{\cK}{{\ml K}}
\nwc{\cB}{{\ml B}}
\nwc{\cD}{{\ml D}}
\nwc{\cF}{{\ml F}}
\nwc{\cM}{{\ml M}}
\nwc{\cG}{{\ml G}}
\nwc{\cH}{{\ml H}}
\nwc{\bk}{{\mb k}}
\nwc{\cQ}{{\ml Q}}
\nwc{\cO}{{\ml O}}
\nwc{\cJ}{{\ml J}}

\nwc{\mint}{{\int\cdot\int}}

\begin{document}

\title{  Nonlinear Schr\"odinger Equation with
a White-Noise Potential: Phase-space Approach to Spread and
Singularity}

\author{Albert C. Fannjiang}
 \email{
  cafannjiang@ucdavis.edu}
 \thanks{Research partially supported by
the Centennial Fellowship from the American
Mathematical Society,
the UC Davis Chancellor's Fellowship
and  U.S. National Science Foundation grant DMS 0306659.
}
 \address{
Department of Mathematics,
University of California, Davis 95616-8633}
\begin{abstract}
We propose a phase-space formulation for
the nonlinear Schr\"odinger equation with
a white-noise potential in order to shed light on two issues: the rate of
spread and the singularity formation in the average sense.
Our main tools are  the energy
law
and  the variance identity. The method is completely elementary.

For the problem of wave spread, we
show that 
 the ensemble-averaged dispersion in the critical or
defocusing case
 follows  the cubic-in-time law while
 in the supercritical and subcritical focusing
 cases the cubic law becomes an upper
 and lower bounds respectively.

We have also found that in the critical and supercritical focusing cases the presence of
a white-noise random potential 
results in different conditions for  singularity-with-positive-probability from the homogeneous case  but
does not prevent singularity formation.
We show that 
in the supercritical focusing case 
 the ensemble-averaged self-interaction energy
and the momentum variance can exceed 
any fixed level in a finite time with positive probability.

\end{abstract}

\maketitle

\section{Introduction}

High-intensity  laser beams propagating in dielectric media
such as optical fibers, films or air are important problems both from fundamental and practical perspectives. 
The physical process is nonlinear and the
amplitude modulation $\Psi$ is customarily described 
by a nonlinear Schr\"odinger equation with
an additional  inhomogeneous  potential $V$ representing
the random impurities in the medium.

Let $z$ and $\bx\in \IR^d$ be the longitudinal and transverse
coordinates of the wave beam. In the physical setting,
the transverse dimension $d$ may be $0,1,2$.
 The simplest non-dimensional form  of the nonlinear Schr\"odinger (NLS)
equation with a random potential $V$
then reads \cite{NM}
\beq
\label{para2}
\label{nls}
i\partz
\Psi+\frac{\gamma}{2}\Delta_\bx\Psi
+\gamma^{-1}g|\Psi|^{2\sigma}\Psi+
\gamma^{-1} V \circ\Psi &=&0,\quad\sigma>0.
\eeq
Similar equations also arise in many other contexts
such as  in plasma physics and the Bose-Einstein condensation,
see \cite{SS} and references therein.
Here we have written the equation in the
comoving  reference frame at
 the group velocity  and non-dimensionalized
 the equation with the longitudinal and
 transverse reference lengths $L_z, L_x$; $\gamma=L_z/(kL_x^2)$ is the Fresnel number;
 $g$ is nonlinear coupling coefficient with
$g>0$ representing the self-focusing case
and $g<0$ the self-defocusing case. 
For a nonlinear Kerr medium $\sigma=1$ leading
to the cubic NLS equation. We shall let $\sigma$ be an arbitrary positive constant.  We shall consider 
white-noise-in-$z$, $\bx$-homogeneous random potential 
$V$ with
\[
\lan V(z,\bx) V(z', \bx')\ran
=\delta(z-z') \int e^{i(\bx-\bx')\cdot\bp}
\Phi(\bp)d\bp
\]
where $\Phi(\bp)$ is the  power spectral density and  the product $V\circ \Psi$ in eq. (\ref{para2}) 
stands for the Stratonovich product. 
Here and below the bracket $\lan \cdot \ran$ denotes
the ensemble average w.r.t. the random medium. In view of the real-valuedness of $V$  we may  assume
\beq
\label{sym}
\Phi(-\bp)
=\Phi(\bp),\quad\forall \bp\in\IR^d
\eeq

Such a short-range-in-$z$ potential arises in long-distance propagation when the longitudinal
length $L_z$ of the wave beam is much larger
than the correlation length of the random impurities
in the medium  resulting in rapidly fluctuating potential
in the non-dimensionalized coordinates. And it is well known
that in the short-range-correlation scaling limit a non-white-in-$z$ 
multiplicative noise gives rise to a  Stratonovich, instead of
It\^o, integral with respect to a white-in-$z$ noise \cite{PK}.

We are particularly
interested in the problems of wave spread and singularity
formation.
To this end we shall use the phase-space approach
of the Wigner distribution which is particularly useful
 in  the regime of low Fresnel number
$\gamma\ll 1$ which can be viewed as 
a high-frequency limit.
The main
ingredient of our analysis is the phase-space
variance identity by which we derive
various estimates for the mean spread, defined in
(\ref{vx}), including precise behavior in
the defocusing $g\leq 0$ or  critical case $d\sigma=2$. 
To our knowledge, these 
are significant improvement over
previous results (e.g.
\cite{BL}, \cite{Tch},  \cite{EKS}) which
are mostly for the linear problem. Wave spread 
is an important physical quantity in itself 
and in technological applications such as 
estimating the effect of nonlinearity on  the information capacity
of optical fiber communications \cite{Kaz}, \cite{Mit}.

Using the variance identity we further 
show that the scattering by the random potentials 
results in  conditions for singularity-with-positive-probability
that are different from those for the homogeneous case.

In the present setting the physical roles of $t$ and $z$ are
interchanged:    the variable $t$ is space-like
while
the variable $z$ is time-like
and 
we will refer to it as ``time'' occasionally, especially
in the discussion of finite-time singularity.

\section{Wigner distribution} 
We consider
the Wigner distribution of the form
\beq
\label{pure}
 W(z,\bx,\bp) =   \frac{1}{(2\pi)^d}
           \int  e^{-i \bp \cdot \by} 
         \Psi\lt(z,\bx+\frac{\gamma \by}{2}\rt)
		{\Psi^*\lt(z, \bx-\frac{\gamma \by}{2}\rt)}
d\by 
\eeq
\commentout{
\beq
\label{pure}
 W(t,\omega, \bx,\bp) =   \frac{1}{(2\pi)^d}
           \int e^{i\omega \tau } e^{-i \bp \cdot \by} 
         \Psi\lt(t+\frac{\delta}{2}\tau, \bx+\frac{\gamma \by}{2}\rt)
		{\Psi^*\lt(t-\frac{\delta}{2}\tau, \bx-\frac{\gamma \by}{2}\rt)}
d\by d\tau
\eeq
}
which is always real-valued and  may be thought of as quasi-probability density on
the phase space \cite{LP}, \cite{GMMP}.  First its marginals
are essentially the position and momentum densities
\beqn
\int W(z,\bx,\bp)d\bp&=& |\Psi|^2(z, \bx)\equiv \rho(z,\bx)\\
\frac{1}{(2\pi)^d}\int W(z,\bx,\bp)d\bx &=& \frac{1}{\gamma^d}
|\hat\Psi(z,\bp)|^2
\eeqn
which are nonnegative;
indeed any marginal on a $d$-dimensional subspace
is nonnegative everywhere. Also, the energy flux is given by
\beq\label{2.2.3}
\frac 1{2i}(\Psi\nabla\Psi^*-\Psi^*\nabla\Psi)=
\int_{\IR^d}\bp W(z, \bx,\bp)d\bp.
\eeq
In the whole phase space, however, the Wigner distribution is not everywhere non-negative in general and hence can not be
a true
 probability density. As $\gamma\to 0$, the Wigner distribution has a nonnegative-measure-valued  weak limit,
 called the Wigner measure \cite{LP}.
Therefore   the Wigner distribution is particularly useful for
 analyzing high frequency behaviors of waves.

 \section{The evolution equations}

From the NLS eq. (\ref{nls})  it is straightforward to derive  the
 closed-form (Wigner-Moyal) equation for the Wigner distribution
 in the It\^o sense \cite{whn-wig}, \cite{rad-arma}
\beq
\label{nwm}
{d_z W}
+{\bp}\cdot\gradx W{dz}+\cU_\gamma Wdz-
\cQ_\gamma Wdz
+\cV_\gamma Wdz=0\label{wig}.
\eeq
Here  the {\em  self-adjoint} operator $\cQ_\gamma$ is the Stratonovich correction term 
\beqn
\lefteqn{{\cQ}_\gamma W(z,\bx,\bp)}\\
&=&
\int \Phi(\bq)
 \gamma^{-2}\lt[-2 W(z, \bx,\bp)+
  W(z, \bx,\bp-\gamma\bq)
  + W(z, \bx,\bp+\gamma\bq)\rt] {d\bq}
  \label{102}
  \eeqn
and $\cU_\gamma$ and $\cV_\gamma $ are
the nonlinear and linear  Moyal operators, respectively
\beqn
\label{cu}
\lefteqn{\cU_\gamma W(z,  \bx,\bp)}\\&=&
i\int e^{i\bq\cdot\bx}
\gamma^{-1}\lt[W(z,\bx,\bp+{\gamma}\bq/2)-W(z,\bx,\bp-
{\gamma}\bq/2)\rt]
\widehat{U}(z,\bq)d\bq,\\
\lefteqn{\cV_\gamma W(z, \bx,\bp)}\\&=&
i\int e^{i\bq\cdot\bx}
\gamma^{-1}\lt[W(z, \bx,\bp+\gamma\bq/2)-W(z, \bx,\bp-
\gamma \bq/2)\rt]
\widehat{V}(d{z},d\bq)
\label{cv}
\eeqn
with
\[
U= g\rho^{\sigma}.
\]
Here we have somewhat abused the notation by writing
$\hat{U}$ as the Fourier-transform-in-$\bx$ of the function
$U(z,\cdot)$ and $ \hat{V}(dz,d\bq)$ as the
spectral measure of the white-noise potential $V(z,\cdot)$.
The spectral measure $\hat{V}(dz,d\bp)$ is related to
the power spectral density $\Phi$ of $V$ as
\[
\lan \hat{V}(dz,d\bp)\hat{V}^*(dz',d\bq)\ran
=\delta(z-z') \delta(\bp-\bq) \Phi(\bp)dz d\bp d\bq.
\]
An important property of these integral operators is
\beq
\label{null}
\int \cQ_\gamma Wd\bp=\int \cU_\gamma Wd\bp=
\int \cV_\gamma W d\bp=0
\eeq
which will be useful in deriving the energy law and the variance
identity. 

A major advantage of formulating the wave process
on the phase space using the Wigner distribution is
that the (high-frequency)  low-Fresnel number limit
$\gamma$ can be easily obtained.
Formally we see that  as
$\gamma\to 0$
\beqn
\label{go}
\cU_\gamma W(z,\bx,\bp)&\to \cU_0W(z, \bx,\bp)&\equiv \gradx
U(z,\bx)\cdot\nabla_\bp W(z, \bx,\bp)\\
\cV_\gamma W(z,\bx,\bp)&\to\cV_0W(z, \bx,\bp)&\equiv \gradx
V(z,\bx)\cdot\nabla_\bp W(z, \bx,\bp)\\
\cQ_\gamma W(z,\bx,\bp)&\to 
\cQ_0W(z,\bx,\bp)&=\nabla_\bp\cdot\bD\nabla_\bp W(z,\bx,\bp)
\eeqn
with the diffusion coefficient
\beq
\label{D1}
\bD=\int\Phi(\bp) \bp\otimes\bp d\bp.
\eeq
\commentout{
We use
the following definition of the Fourier transform 
and inversion:
\beqn
\cF f(\bp)&=&\frac{1}{(2\pi)^d}\int e^{-i\bx\cdot\bp}
f(\bx)d\bx\\
\cF^{-1} g(\bx)&=&\int e^{i\bp\cdot\bx} g(\bp) d\bp.
\eeqn
We only need to  consider the weak formulation
\beq
\partz \int \Theta Wd\bx d\bp
-\int \bp\cdot\gradx \Theta Wd\bx d\bp
-\int \cU_\gamma \Theta Wd\bx d\bp
-\mu \int \cV \Theta W d\bx d\bp
=0
\label{nwm}
\eeq
for smooth, rapidly decaying test functions
$\Theta$.
}
We shall refer to eq. (\ref{nwm}) 
as the nonlinear Wigner-Moyal -It\^o (NWMI) equation when
$\gamma>0$
and as the nonlinear Liouville-It\^o (NLI) equation
when $\gamma=0$.

Another advantage of working with eq. (\ref{nwm})
is that one can use it to evolve
the mixed-state initial condition. 
The mixed-state Wigner distribution is
a convex  combination of the pure-state
Wigner distributions (\ref{pure}) described as follows.

Let $\{\Psi_\alpha\}$ be a family of  $L^2$ functions
parametrized by $\alpha$ which is weighted by
a probability measure $P(d\alpha)$ .
 Denote the pure-state Wigner distribution
(\ref{pure}) by $W[\Psi_\alpha]$. A mixed-state Wigner
distribution is given by
\beq
\label{mixed}
\int W[\Psi_\alpha]P(d\alpha).
\eeq
The limiting set, as $\gamma\to 0$, of the mixed state
Wigner distributions constitute the nonnegative
Wigner measures 
 \cite{GMMP}, \cite{LP}. The NWMI
or NLI equation preserves the mixed-state form (\ref{mixed}).
\commentout{
 In
particular, for such initial data
we have
\beq
\label{pos1}
\int W(z,\bx,\bp)d\bp&\geq& 0,\quad \forall \bx \in
\IR^d\\
\int W(z,\bx,\bp)d\bx &\geq& 0,\quad \forall \bp\in
\IR^d
\label{pos2}
\eeq
Multiplying (\ref{nwm}) by $W$ and integrating
by parts we also see that the evolution conserves
the $L^2(\IR^{2d})$-norm of $W$, i.e.
\[
\partz \int | W|^2 d\bx d\bp =0.
\]
}

\subsection{Basic properties}
The NWMI and NLI equations  conserve
the total mass, i.e.
\[
\partz N=0,\quad N= \int W(z,\bx,\bp)d\bx d\bp
\]
and the $L^2$ norm
\[
\partz \int W^2(z,\bx,\bp)d\bx d\bp=0.
\]
One
can absorb the effect of the total mass $N$ into
$g$ by the obvious rescaling of $W$ in the eq.
(\ref{nwm}). Henceforth we assume
that $
N=1$.

Let $S_x$ and $S_p$ be the spreads (or variances) of  position and momentum, respectively
\beq 
S_x(z)&=&\int |\bx|^2 W(z, \bx,\bp)d\bx d\bp\label{vx}\\
S_p(z)&=&\int
|\bp|^2W(z, \bx,\bp)d\bx d\bp.
\eeq
A natural space of initial data and solutions is
the subspace $\cW\subset L^2(\IR^{2d})$ with a finite Dirichlet form
\[
-\int W\cQ_\gamma Wd\bx d\bp<\infty
\]
and finite, positive (pre-ensemble-averaged) variances
$S_x\in 
(0,\infty),$$
S_p\in (0,\infty).$
In addition, we shall also assume that
the initial data have a finite
 Hamiltonian  $H\in (-\infty,\infty)$ with 
\beq
\label{Ham}
H=\frac{1}{2}S_p-\frac{g}{\sigma+1}
\int \rho^{\sigma+1}d\bx, \quad
\rho(z,\bx)=\int W(z,\bx,\bp)d\bp.
\eeq
The first term $S_p$ in $H$ is the kinetic energy
and the second term is  the (self-interaction) potential energy.
Note that in the presence of a random potential,
the value of the Hamiltonian (\ref{Ham}) is not conserved
under the evolution (\ref{nwm}). Indeed, the  ensemble-averaged value of
the Hamiltonian increases in time, cf. (\ref{H2}) below.

\section{Energy and variance identity}
Let us analyze 
the evolution of the mean Hamiltonian $\lan H\ran$ and the
average spread
$\lan S_x\ran$.  First note the result of $\cU_\gamma$ 
acting on  the quadratic polynomials: 
\beq
\label{ga1}
\cU_\gamma \bx&=&0\\
\cU_\gamma \bp&=&i\int e^{i\bq \cdot \bx}\bq
\hat{U}(\bq)d\bq=\gradx U\\
\cU_\gamma |\bx|^2&=&0\\
\cU_\gamma \bx\cdot\bp&=&i\int e^{i\bq\cdot\bx}\bx\cdot\bq
\hat{U}(\bq)d\bq=\bx\cdot\gradx U\\
\cU_\gamma |\bp|^2&=& i\int e^{i\bq\cdot\bx}
2\bp\cdot\bq \hat{U}(\bq)d\bq=2\bp\cdot\gradx U.
\label{ga2}
\eeq
A remarkable fact is that these results 
are 
independent of $\gamma\geq 0$.

Using (\ref{ga1})-(\ref{ga2}) and (\ref{null})  we obtain
by integrating eq. (\ref{nwm}) that
\beqn
\partz \lan S_p\ran &=&\lan \int \nabla_\bx U \cdot 2\bp Wd\bx d\bp\ran
+\int \cQ_\gamma|\bp|^2 \lan W\ran d\bp d\bx\\
\partz \frac{g}{\sigma+1}\int \lan  \rho^{\sigma+1}\ran d\bx&=&
\lan \int \nabla_\bx U\cdot \bp Wd \bp d\bx\ran.
\eeqn
from which the
evolution equation for the mean value of
the  Hamiltonian follows
\beq
\partz \lan H\ran &=&\int \cQ_\gamma|\bp|^2 \lan W\ran d\bp d\bx.
\label{h2}
\eeq
We shall refer to (\ref{h2})  as the {\em energy law}.

\subsection{Variance identity}
The variance identity has been long used to
derive the wave collapse condition for 
the NLS in the homogeneous case \cite{SS}.
Below we reformulate it for the phase space 
evolution eq. (\ref{nwm}).

We have another set of remarkably simple properties
of
$\cQ_\gamma$ independent
of $\gamma\geq 0$:
\beq
\label{q1}
\cQ_\gamma \bx&=& 0\\
\cQ_\gamma\bp&=&\int \Phi(\bp-\bq)(\bp-\bq)d\bq=0\quad
(\hbox{by (\ref{sym})})\\
\cQ_\gamma |\bx|^2&=&0\\
\cQ_\gamma \bx\cdot\bp&=&
\bx\cdot \int \Phi(\bp-\bq)(\bp-\bq)d\bq=0\quad
(\hbox{by (\ref{sym})})\\
\cQ_\gamma |\bp|^2&=&-\int \Phi(\bp-\bq)
\lt[ |\bp|^2-|\bq|^2\rt] d\bq \\
&=& -\int \Phi(\bq)(2\bp-\bq)\cdot\bq
d\bq\\
&=& \int \Phi(\bq)|\bq|^2d\bq\equiv  2R.
\label{q2}
\eeq
For the diffusion operator $\cQ_0$ we have 
$
R=2\times\hbox{tr}(\bD)
$
where the diffusion matrix $\bD$ is given
by (\ref{D1}).
We shall use the above identities
to perform integrating by parts in
the derivation of the variance identity.

Combining  the results from the previous section and
(\ref{q1})-(\ref{q2})
we obtain the rate of change of
$\lan S_x\ran$
\beq
\label{blow}
\partz \lan S_x\ran
&=&2 \lan S_{xp}\ran 
\eeq
where $S_{xp}$ is the cross-moment 
\[
S_{xp}=\int \bx\cdot\bp Wd\bx d\bp.
\]
Differentiating $S_{xp}$ and taking expectation we obtain
\beqn
\partz \lan S_{xp}\ran 
\nn&=&\lan S_p\ran -\frac{gd\sigma}{\sigma+1}\int 
\lan \rho^{\sigma+1}\ran d\bx d\bp
\eeqn
Hence the second derivative of $S_x$ becomes
\beq
\frac{\partial ^2}{\partial z^2} \lan S_x\ran 
&=&4\lan H\ran  +\frac{2(2-d\sigma)g}{\sigma+1}\int
\lan \rho^{\sigma+1}\ran d\bx.
\label{var1}
\eeq
Alternatively using the definition of $H$ we can
rewrite the variance identity (\ref{var1}) as
\beq
\label{var2}
\frac{\partial ^2}{\partial z^2} \lan S_x\ran 
 &=&2d\sigma\lan H\ran +(2-d\sigma)\lan S_p\ran
\eeq
which in the critical case $d\sigma=2$ becomes
\beq
\label{var3}
\frac{\partial ^2}{\partial z^2} \lan S_x\ran 
 &=&2d\sigma\lan H\ran .
\eeq
Both (\ref{var1}) and (\ref{var2}) will
be useful for estimating wave spread.

\section{Dispersion rate}

Although the medium is lossless, reflected in
the fact that the total mass $N=1$ is conserved,
 the value of the Hamiltonian, however,  is not conserved by the evolution
since the random scattering is
not elastic due to time-varying nature of the random potential.
Indeed, by (\ref{h2}) and (\ref{q2}), the average Hamiltonian is an increasing function of time
\beq
\label{H2}
\partz \lan H\ran &=& R,\quad \lan H\ran (z)=H(0) +Rz
\eeq
 due to the diffusion-like spread in the momentum
$\bp$.

In the critical case $d\sigma=2$, we obtain from
eq. (\ref{var3}) and (\ref{H2})
the exact result \beq
\label{31'}
\frac{\partial^2}{\partial z^2}
\lan S_x\ran =4\lan H\ran=4 H(0)+4R z
\eeq
before any singularity formation causes a possible
breakdown of the variance identity.
Integrating (\ref{31'}) twice we obtain the exact spread
rate as stated below. 
\begin{prop} 
If $d\sigma=2$ or $g=0$, then
\beq
\label{exact1}
\lan S_x\ran (z)=S_x(0) +2S_{xp}(0)z+2H(0)z^2+
\frac{2R}{3} z^3.
\eeq

\end{prop}
The analogous result ($S_x\sim z^3$) for the 
linear  Schr\"odinger equation ($d=1, g=0$)
with a random potential has been proved
previously \cite{BL}, \cite{EKS}. 

Next we consider the supercritical case and the defocusing
case.
We have from (\ref{var1})
that
\beq\label{var11}
\frac{\partial ^2}{\partial z^2}
\lan S_x\ran &=&4 \lan H\ran +\frac{(4-2d\sigma)g}{\sigma+1}
\int\lan  \rho^{\sigma+1}\ran d\bx
\eeq
and hence 
\beqn
\frac{\partial ^2}{\partial z^2}
\lan S_x\ran &\leq &4
\lan H\ran=4H(0)+4Rz,\quad\hbox{for}\,\,g(2-d\sigma
)<0\\
\frac{\partial ^2}{\partial z^2}
\lan S_x\ran &\geq &4
\lan H\ran=4 H(0)+4Rz,\quad\hbox{for}\,\,g(2-
d\sigma)\geq
0.
\eeqn
On the other hand, from (\ref{var2})
we obtain 
for any $g$
\beqn
\frac{\partial ^2}{\partial z^2} \lan S_x\ran
&\leq&2d\sigma \lan H\ran,\quad
\hbox{for}\,\, 2-d\sigma\leq 0\\
\frac{\partial ^2}{\partial z^2}\lan  S_x\ran
&\geq&2d\sigma \lan H\ran,\quad
\hbox{for}\,\, 2-d\sigma\geq 0.
\eeqn
Integrating the above inequalities twice, we obtain
the following.
\begin{prop}
\label{prop2}
The following estimates hold
\beq
\label{up2}
&&\hspace{-1cm}\lan S_x(z)\ran\leq S_x(0)+2S_{xp}(0) z+2 H(0)z^2
+\frac{2}{3}Rz^3,\quad g(2-d\sigma)\leq
0\\
&&\hspace{-1cm}\lan S_x(z)\ran \geq  S_x(0)+2S_{xp}(0)z+2H(0)z^2
+\frac{2}{3}Rz^3,\quad g(2-d\sigma)\geq
0 
\eeq
and
\beq
\label{up1}
\lan S_x(z)\ran &\leq& S_x(0) +2S_{xp}(0)z+d\sigma H(0)z^2
+\frac{d\sigma}{3}Rz^3,\quad 2\leq d\sigma\label{45}\\
\lan S_x(z)\ran&\geq& S_x(0) +2S_{xp}(0)z+d\sigma H(0)z^2
+\frac{d\sigma}{3}Rz^3,\quad 2\geq d\sigma.
\label{dn1}
\eeq

\end{prop}

Therefore
\begin{cor}
Assume $ g< 0$ (hence $H\geq 0$). Then
\beqn
 \lan S_x(z)\ran \leq
S_x(0)+2S_{xp}(0)z+(d\sigma\vee 2) H(0) z^2
+\frac{d\sigma \vee 2}{3}Rz^3
\eeqn
and
\beq
\label{defocus1}
 \lan S_x(z)\ran \geq
S_x(0)+2S_{xp}(0)z+(d\sigma\wedge 2) H(0)z^2
+\frac{d\sigma\wedge 2}{3}Rz^3.
\eeq
\end{cor}
In other words the variance $\lan S_x\ran$ is cubic-in-time
in the defocusing case. In the subcritical  case
the cubic law is a lower bound while in the supercritical
case the cubic law
is an upper bound.

\section{Finite-time singularity}
\label{sing-sec1}
Finite-time singularity for the
critical or supercritical ($d\sigma \geq 2$) self-focusing NLS 
equation
in the homogeneous case is a well known effect
\cite{SS}. In this case the singularity is the
blow-up type $S_p, \int|\rho|^{\sigma+1}
\to \infty$.  Here we  call breakdown of
the variance identity (\ref{var1})-(\ref{var2}) or
the energy law (\ref{h2}) as finite-time singularity and seek the
 sufficient conditions for singularity with positive probability.
We show that in the supercritical case with additional assumptions
the finite-time singularity is of the blow-up
type in the sense that $\lan S_p\ran, 
\int \lan \rho^{\sigma+1}\ran$ tend to infinity.
As such  the blow-up phenomenon discussed here is
not necessarily a sure  event but rather an event of a positive
probability that $S_p$ and $\int \rho^{\sigma+1}$ can exceed  any
fixed level in a finite time.

For
$g\geq 0, d\sigma
\geq 2$  one can bound  
$ S_x$ as in the  inequality (\ref{up1}) \beq
\label{vin}
\label{vin1}
\lan S_x(z)\ran \leq S_x(0)+2S_{xp}(0) z+d\sigma H(0) z^2
+\frac{d\sigma R}{3} z^3\equiv F(z)
\eeq
and as motivated by the homogeneous case we look for the conditions when
$F(z)$ vanishes at a finite $z\geq 0$.

A sufficient condition for $F(z)$ to vanish
at a finite positive $z$ can be derived from
 that $F(z)$ takes
a non-positive value  $F(\zo)\leq 0$ at its
 local minimum point $\zo>0$.
The local
minimum point
$\zo$  is given by
\beq
\label{zo}
\zo=\frac{ - H(0)+
\sqrt{H(0)^2-
2R S_{xp}(0)/(d\sigma)}}{ R}.
\eeq
Therefore we are led to the following conditions for
singularity.
\begin{prop}
\label{sing-prop1} 
Assume $d\sigma
\geq 2,g> 0$. 
The solutions of the NWMI or NLI equation
develop singularities with positive probability  at a finite time $z_*\leq \zo$
given by (\ref{zo}) under the condition $F(\zo)\leq
0$ and either one of the following conditions
\beq
\label{in} S_{xp}(0)&<&0\\
\label{out}  S_{xp}(0) &>&0,\quad 
H(0)<-\sqrt{\frac{2R S_{xp}(0)}{d\sigma}}.\eeq
\end{prop}
\begin{remark}
Clearly, the condition $F(\zo)\leq 0$ requires
$ H(0)$ to be
sufficiently below $ \Hbar $ by allowing
the self-interaction  energy 
\[
-\frac{g}{\sigma+1}\int \rho^{\sigma+1}(0)d\bx d\bp
\]
to be sufficiently negative.
\end{remark}

\begin{remark}
Using, instead, inequality (\ref{up2}) one can obtain
an alternative expression
\beq
\label{zo2}
\zo=\frac{- H(0)+
\sqrt{ H(0)^2-
R S_{xp}(0)}}{ R}
\eeq
and the corresponding conditions for singularity formation,
namely
either $
 S_{xp}(0)<0$ or $
 S_{xp}(0) >0, 
H(0)<-\sqrt{R S_{xp}(0)}.$
\end{remark}

 It may be the case
 that for a given initial condition  the solutions develop
 singularity at different times
 depending on the realization of random potential.
 To analyze such an effect we need
 the probabilistic  versions of
 variance identity and energy law which are
 much more involved.


\commentout{
In the linear or self-defocusing case $g\leq 0$ the
right side of (\ref{exact1}) or (\ref{defocus1}) is
always positive in view of the inequality
\beqn
2|\lan V_{xp}\ran |\leq \lan V_x\ran +\lan V_p\ran,
\eeqn
cf. (\ref{cs}).

Clearly the
above result implies that the random potential does not
prevent large scale singularity.

With the assumption that the variance
identity and the energy law 
(\ref{h2}) hold for the maximally extended local
solutions of eq. (\ref{nwm}) then it is clear that the
singularity is the blow-up type
 \[
 \lim_{z\to z_*}\int \lan \rho^{\sigma+1}\ran d\bx=\infty
 \]
 or equivalently
 \[
 \lim_{z\to z_*}\lan S_p\ran =\infty
\]
by the finitude of the Hamiltonian.
}
\subsection{Blow-up}
We will follow the argument of
\cite{Gl} to show  more explicitly the
blow-up mechanism  in the case  with
the supercritical, self-focusing
nonlinearity  and give a sharper bound
on $z_*$ under certain circumstances. 

\begin{prop}\label{blow1}
Suppose 
$d\sigma>2, g> 0$. Then under the
conditions
(\ref{in}) and 
\beq
\label{time2}
\lan H(z_*)\ran =H(0)+R z_*\leq 0,\quad
\hbox{with}\,\,z_*=\frac{2S_x(0)}{S_{xp}(0)(2-d\sigma
)},
\eeq
$\partial \lan S_x\ran/\partial z$ and $\lan S_p\ran$ blow up 
before  or at a finite time
$z_*$.
\end{prop}
{\bf Proof.}
Since blow-up is a local phenomenon, $S_x$ is a poor
indicator of  its occurrence. 
A more relevant object to consider  is $S_p$.
From (\ref{var2}) it follows
that before $z_*$
\beq
\label{var2.1}
\frac{\partial^2}{\partial z^2}\lan  S_x\ran
=2\frac{\partial}{\partial z}\lan  S_{xp}\ran
&\leq &(2-d\sigma)\lan S_p\ran<0.
\eeq
Since $\partial\lan S_x\ran /\partial z=2\lan  S_{xp}\ran<0$ by
(\ref{var2.1}) and (\ref{in}),
$\partial\lan S_x\ran /\partial z$ is a negative, decreasing function
up to the time $z_*$. Hence we have 
\beq
\label{4.4}
0\leq \lan S_{xp}\ran^2\leq
\lan S_x\ran\lan S_p\ran\leq {S}_x(0)\lan S_p\ran,\quad  z\in [0,z_*]
\eeq
which implies 
\beq
\label{47}
\lan S_p\ran \geq \frac{\lan S_{xp}\ran^2}{S_x(0)}.
\eeq

Let $A(z)=-\lan \partial S_x(z)/\partial z\ran \geq 0, z\leq z_*$. We have from
(\ref{var2.1}), (\ref{47}) and (\ref{blow}) the differential
inequality
\beq
\label{grow}
\partz A\geq CA^2,\quad C=\frac{d\sigma-2}{4{S}_x(0)}>0
\eeq
which yields the estimate
\[
A(z)\geq \frac{A(0)}{1-zCA(0)},\quad
z< \frac{1}{CA(0)}=z_*
\]
and thus the blow-up of
$A(z) $ before or at $z_*$. 
This along with (\ref{47}) then 
implies the blow-up of $S_p$  at a finite time.
$\Box$

The preceding argument 
demonstrates clearly the blow-up mechanism, namely
the quadratic growth property (\ref{grow}).
\begin{remark}
It should be noted that, since $H(z_0)>0$, the condition (\ref{time2}) implies
that $z_*<z_0$, as given by either (\ref{zo}) or (\ref{zo2}).
Therefore, under the condition (\ref{time2}), 
$z_*$ provides a sharper upper bound on the time
of singularity than $z_0$. 
\end{remark}
\begin{remark}
Since  $\lan H(z)\ran$ is bounded over compact
sets of $z$, Proposition~\ref{blow1} implies
the blow-up of the self-interaction energy $\int \lan \rho^{\sigma+1}\ran$
in a finite time.
\end{remark}

\section{Conclusion}
We have presented an elementary approach for
analyzing the nonlinear Schr\"odinger equation
with a white-noise potential. We have focused
on the ensemble-averaged quantities such
as the variance identity and energy law
in order to shed light on two problems: the rate of
spread and the singularity formation.

We have 
shown that 
the ensemble-averaged spread in the critical or
defocusing case
 follows  the cubic-in-time law while
 in the supercritical and subcritical focusing
 cases the cubic law becomes an upper
 and lower bounds respectively. In a separate
 publication we will use these estimates to
 analyze the limitation on channel capacity in optical
 fibers due to self-phase and cross-phase modulations
 \cite{Kaz}. 

We have also found singularity conditions in
the critical and supercritical focusing cases.
And we show the finite-time singularity in the supercritical
case is of the blow-up type. The singularity/blow-up discussed
in the present paper is not necessarily a sure event
but that of a positive probability.

\end{document}